\documentclass{article}
\usepackage[english]{babel}
\usepackage{amsmath,amsthm,eepic,color}
\usepackage{amsfonts}
\usepackage{graphicx}
\usepackage{bm}
\newtheorem{remark}{{Remark }}

\def\openone{\leavevmode\hbox{\small1\kern-3.3pt\normalsize1}}

\def\diag{\mbox{diag\,}}

\def\const{\mbox{const\,}}
\def\ad{\mbox{ad\,}}

\title{Recursion operators and the hierarchies of MKdV equations related to $D_4^{(1)}$,  $D_4^{(2)}$  and $D_4^{(3)}$
Kac-Moody algebras}
\author{V. S. Gerdjikov$^{1,2,3}$, A.A. Stefanov$^{1,4}$,  I. D. Iliev$^1$,  G. P. Boyadjiev$^1$, \\
A. O. Smirnov,$^{5}$  V. B. Matveev,$^{6,7}$ M. V. Pavlov$^{8}$\\
  \\[5pt]
{\sl $^1$ Institute of Mathematics and Informatics,  Bulgarian Academy of Sciences, }\\
{\sl Acad. Georgi Bonchev Str., Block 8, 1113 Sofia, Bulgaria}\\[5pt]
{\sl $^2$ National Research Nuclear University MEPHI, }\\
{\sl 31 Kashirskoe Shosse, 115409 Moscow, Russian Federation}\\[5pt]
{\sl  $^3$ Institute for Advanced Physical Studies,  New Bulgarian University,}\\
{\sl 21 Montevideo Street, Sofia 1618, Bulgaria} \\[5pt]
{\sl $^4$ Faculty of Mathematics and Informatics, Sofia University "St. Kliment Ohridski"} \\
{ 5 James Bourchier Blvd.,  1164 Sofia, Bulgaria } \\[5pt]
{$^5 $\sl Sankt-Petersburg    State    University    of   Aerospace
Instrumentation } \\ {\sl   St-Petersburg,   B.Morskaya,   67A,  St-Petersburg, 1900000, Russia}\\[5pt]
{$^6$ \sl Sankt-Petersburg department of Steklov Mathematical Institute} \\
{\sl of Russian Academy of Sciences,St-Petersburg, Russia}\\[5pt]
{$^7$  \sl Institut de Math\'ematiques de Bourgogne (IMB),} \\
{\sl Universit\'e de Bourgogne - France Comt\'e, Dijon, France} \\[5pt]
{\sl $^8$  P.N. Lebedev Physical Institute of Russian Academy of Sciences} \\
{\sl Leninskij Prospekt, 53, Moscow, 119991, Russia.} \\
 }

\date{ }

\begin{document}
\maketitle
\thispagestyle{empty}
\begin{abstract}
We constructed the three nonequivalent gradings in the algebra $D_4 \simeq so(8)$. The first one is the
standard one obtained with the Coxeter automorphism $C_1=S_{\alpha_2} S_{\alpha_1}S_{\alpha_3}S_{\alpha_4}$ using
its dihedral realization. In the second one we use $C_2 = C_1R$ where $R$ is the mirror automorphism. The third one
is $C_3 = S_{\alpha_2}S_{\alpha_1}T$ where $T$ is the external automorphism of order 3. For each of these gradings we
constructed the basis in the corresponding linear subspaces $\mathfrak{g}^{(k)}$, the orbits of the Coxeter automorphisms
and the related Lax pairs generating the corresponding mKdV hierarchies. We found compact expressions for each of the hierarchies
in terms of the recursion operators. At the end we wrote explicitly the first nontrivial mKdV equations and their Hamiltonians.
For $D_4^{(1)}$ these are in fact two mKdV systems, due to the fact that in this case the exponent $3$ has multiplicity 2.
Each of these mKdV systems consist of 4 equations of third order with respect to $\partial_x$. For $D_4^{(2)}$ this is a
system of three equations of third order with respect to $\partial_x$. Finally, for $D_4^{(3)}$ this is a system of two
equations of fifth order with respect to $\partial_x$.
\end{abstract}

\tableofcontents
\pagenumbering{arabic}

\section{Introduction}

\label{intro}

The works of Gardner, Green, Kruskal and Miura \cite{GGKM} and  Lax \cite{Lax} can be considered the foundation of modern soliton science.  Initially many researchers believed that the methods presented in those papers apply only to the Korteweg-de-Vries (KdV) equation.

The second soliton equation discovered by Zakharov and Shabat in 1971 \cite{ZakharovShabat0} was the nonlinear Schr\"odinger equation (NLS); soon after that the third
integrable equation, the mKdV appeared \cite{Wadati-MKdV}.
It was followed by an explosion of interest in soliton equations. Mathematicians were excited by the fact that both KdV and NLS provided the first examples of
infinite dimensional completely integrable Hamiltonian equations \cite{Zakharov-Faddeev,Takht,Bullough-Caudrey}. The physicists appreciated the new stable nonlinear waves that had
purely elastic interaction and appeared  in various physical processes: hydrodynamics, plasma physics, nonlinear optics etc.

It is worth mentioning some of the milestones in this development.

The seminal paper by Ablowitz, Kaup, Newell and Segur \cite{AKNS} demonstrated new techniques of working out with the Lax pairs
and formulated new important idea. The proved that the inverse scattering method (ISM) can be understood as a generalized
Fourier transform (GFT), which allows one to linearize the soliton equations. They introduced the notion of the recursion operator $\Lambda$,
that generated the hierarchy of soliton equations and the GFT were the spectral expansions of $\Lambda$. The proof of this idea was completed
by establishing the completeness relations for the `squared solutions` of the Lax operator $L$ \cite{Kaup*76, GeKh1, KauNew*78, IlKh}.

The next examples of soliton equations such as the $N$-wave equations \cite{ZaMa}, the principal chiral field \cite{ZaMi} and the massive Thirring
model \cite{KuzMi} quickly led to the necessity to extend the ISM to new classes of Lax operators depending polynomially on
the spectral parameter $\lambda$, see also \cite{AthoFord, Gadj, GI1, GI2, GIK-tmf}. Indeed, the AKNS system provided the simplest
nontrivial Lax operator
\begin{equation}\label{eq:La0}\begin{split}
L\psi \equiv i \frac{\partial \psi}{ \partial x } + (Q(x,t) - \lambda J)\psi(x,t,\lambda)=0
\end{split}\end{equation}
which was linear in the spectral parameter $\lambda$, with $J= \sigma_3$ and whose potential $Q(x,t)$ took values in the algebra $sl(2)$.
We note also that the ISP for this Lax operator has been developed earlier by Gelfand, Levitan and Marchenko \cite{GeLeMa}.
At the same time the $N$-wave equations required Lax operators like (\ref{eq:La0}) but with potentials taking values in the algebra $sl(n)$
and $J$ being real constant diagonal matrix $J =\diag (a_1,a_2,\dots, a_n)$.
The principle chiral fields and the massive Thirring models demonstrated the need to study the
spectral properties of operators, that had more complicated  dependence of $\lambda$: rational for the chiral field and polynomial in $\lambda$
and $\lambda^{-1}$.

The direct and inverse scattering problems (ISP) for the $n\times n$ operators linear in $\lambda$ were solved introducing  the notion of a
fundamental analytic solution of $L$ \cite{Sh*75, Sh*79}. As a result it became clear that the ISP is equivalent to a Riemann-Hilbert problem
(RHP). The next important step here was proposed by Zakharov and Shabat, who developed the dressing method \cite{ZakharovShabat1, ZakharovShabat2, NMPZ}
for constructing the soliton solutions of the relevant soliton equations.
Rather quickly it was demonstrated that the AKNS idea of interpreting the ISM as a GFT can be generalized for Lax operators related
not only to $sl(n)$ \cite{GeKu0}, but also to any simple Lie algebra $\mathfrak{g}$ \cite{VG-IP2}. Lax operators polynomial in $\lambda$ were used to
integrate the so-called derivative NLS equation \cite{KauNew*78} and its gauge equivalent GI equation \cite{GI1, GI2}. And again it was possible to
demonstrate that the ISM is a GFT \cite{GIK-tmf, GI1, GI2}.

Another important class of generalizations applied to the soliton equations  was established by Kulish and Fordy \cite{ForKu}. The discovered the
fact that using symmetric spaces one can construct multicomponent generalizations of the corresponding NLS or GI equations, see \cite{Basic}. Again
one can naturally extend the main tools of the soliton equations like the RHP and the dressing Zakharov-Shabat method for obtaining the soliton solutions.
The notion of `squared solutions` and the ideas that ISP is a GFT are also naturally generalized, see \cite{Basic} and references therein.

Another important question that came up was: given a NLEE can we check if it is integrable or not? A way to answer it was to check whether the equation
possesses an infinite set of integrals of motion, or symmetries. Following this ideas Shabat, Zhiber, Mikhailov \cite{ZhiSha, MiShaYa, MiShaYa1} developed a method for classification of
all integrable NLEE of given form, see \cite{MNW,Novik1} and the references therein. Some of these equations, like the one now known as Tsitseica eq. \cite{Tzitzeica}:
\begin{equation}\label{eq:Tzi}\begin{split}
 u_{xt} = e^{u} - e^{-2u},
\end{split}\end{equation}
became for some time a challenge. It was known to have physical applications \cite{DodBul}, it was  known to have an infinite number of integrals
of motion but its Lax representation for some time was unknown. The reason for that was not only that the relevant Lax operator was related to the $sl(3)$
algebra, but also in the fact that it had very special symmetry. Solving this problem A. V. Mikhailov introduced the so called reduction group \cite{Mikhailov} and
discovered the family of 2-dimensional Toda field theories (see also \cite{MiOlPer}), of which Tsitseica equation was a member. These trend was later
extended to treat generalizations of mKdV equations in \cite{MNW, Mikhailov,  Novik1}.

Solving the ISP for Lax operator of the form (\ref{eq:La0}) with $\mathbb{Z}_h$ reduction group leads to the necessity to consider $J$ with complex-valued
eigenvalues. The construction of the fundamental analytic solutions for this class of Lax operators was achieved by Beals and Coifman \cite{Beals-Coifman} for
systems related to $sl(n)$ algebras. Later their results were generalized to any simple Lie algebras \cite{GeYa*94}. The completeness of the `squared solutions`
and the ideas of GFT \cite{GI1, GI2, GIK-tmf, GeKu0, GeYaV} were combined with the Mikhailov reduction group in \cite{VG-Ya-13, VG-Ya-14}.

 Another important aspect, namely that there is a connection between  soliton equations and Kac-Moody algebras  was  discovered  by  Drifneld  and  Sokolov
 \cite{DriSok1, DriSok}. At that time the Lax pairs for the KdV and mKdV equations were often formulated using scalar
 differential operators of third order. Drinfeld and Sokolov demonstrated that scalar differential operators of order $n$ can be conveniently rewritten
 as first order $n\times n$ operator with conveniently applied $\mathbb{Z}_n$ reduction. They extended this result by showing the deep connection between
 the soliton equations and the Kac-Moody algebras. The latter can be constructed starting from  simple Lie algebra $\mathfrak{g}$ graded by its Coxeter
 automorphism $C$, for details see Section 2 below.

This present paper is an extension of our previous results reported in \cite{GMSV3, GMSV4}.
Its main purpose is to present the modified Korteweg-de-Vries (mKdV) equations related to Kac-Moody algebras of type $D_4^{(k)}$, $k=1, 2, 3$.
We assume that the reader is familiar with the theory of simple Lie algebras \cite{Helgasson} and with the basic ideas for constructing Kac-Moody algebras \cite{Carter}.
In Section 2 we outline the construction of the three nonequivalent gradings of the algebra $D_4 \simeq so(8)$ which give rise
to the three Kac-Moody algebras of height 1, 2 and 3.
In Section 3 we formulate the Lax pairs related to each of the three gradings. Extending the AKNS ideas we solve the recurrent relations for each
of the hierarchies of mKdV equations. To this end we have to introduce several types of elementary recursion operators $\Lambda_a$. We also introduce a master
recursion operator $\Lambda$ which is  an ordered product of the elementary ones. Then we find that with each exponent of the Kac-Moody algebra one can relate
a hierarchy of mKdV equations generated by the master recursion operator $\Lambda$. An exception is the case $D_4^{(1)}$ for which the exponent 3 is double-valued.
As a result with this exponent one can relate two nonequivalent hierarchies of mKdV equations.
In Section 4 we provide explicitly the simplest mKdV equations.  For the case $D_4^{(1)}$ these are two systems of 4 equations of third order with respect to
$\partial_x$. For the case $D_4^{(2)}$ this is a system of three equations  of third order with respect to $\partial_x$. Finally for $D_4^{(3)}$ this is a
system two equations of fifth order with respect to $\partial_x$. We also briefly analyze their relations with the results of \cite{MNW}.
In the last Section we briefly discuss the results and outline their possible extensions.

\section{Kac-Moody  Algebras of $D_4$-type}
We assume that the reader is familiar with the basic facts about the simple Lie algebras \cite{Helgasson, Kac}.
\subsection{Kac-Moody algebras}
Let $\mathfrak{g}$ be a finite-dimensional Lie algebra over $\mathbb{C}$. Then
\begin{equation}
\begin{aligned}
\mathfrak{g}[\lambda, \lambda^{-1}]&=\left \{ \sum_{i=n}^{m} v_i \lambda^i : v_i \in \mathfrak{g} , n,m \in \mathbb{Z} \right \}, \\
f[\lambda] &= \left \{ \sum_{i=0}^{m} f_i \lambda^i : f_i \in \mathfrak{g} , m \in \mathbb{Z} \right \}.
\end{aligned}
\end{equation}
There is a natural Lie algebraic structure on $\mathfrak{g}[\lambda, \lambda^{-1}]$.
Let $\varphi$ be an automorphism of $\mathfrak{g}$ of order $s$. Then
\begin{equation}
L(\mathfrak{g}, \varphi) =  \left \{ f \in \mathfrak{g}[\lambda, \lambda^{-1}]: \varphi(f(\lambda))= f \left[ \lambda \exp \left({\frac{2\pi i}{s}}\right) \right] \right \}.
\end{equation}
$L(\mathfrak{g}, \varphi)$ is a Lie subalgebra of $\mathfrak{g}[\lambda, \lambda^{-1}]$.
If $\mathfrak{g}$ is simple then $ L(\mathfrak{g}, \varphi)$ is called a Kac-Moody algebra.
It is obvious that Kac-Moody algebras are graded algebras. Note that commonly the central extension of $L(\mathfrak{g}, \varphi )$ is called a Kac-Moody algebra.
The definition given above is the one used in \cite{DriSok1, DriSok}.

The above definition of Kac-Moody algebras can be stated in simpler words -  the elements of a Kac-Moody algebras are formal series in $\lambda$ with coefficients in some properly graded finite-dimensional simple Lie algebra.

As is shown in \cite{Kac}, two Kac-Moody algebras $L(\mathfrak{g_1}, \varphi_1), L(\mathfrak{g_2}, \varphi_2)$ are isomorphic if $\mathfrak{g_1}$ is isomorphic to $\mathfrak{g_2}$ and
the automorphisms of the Dynkin diagram determined by $\varphi_1$ and $\varphi_2$ are conjugate. Since there are simple Lie algebras with non-trivial outer automorphisms, for those simple Lie algebras there will be more than one Kac-Moody algebra.
Again, every automorphism $\varphi$ of $\mathfrak{g}$ can be uniquely represented in the form $\varphi = f \circ \varphi_{\tau}$. The order of $\varphi_{\tau}$ is called the height of $L(\mathfrak{g}, \varphi)$. Commonly, Kac-Moody algebras of height greater than one are called twisted Kac-Moody algebras.

 In analogy with finite dimensional simple Lie algebras, $C$ is a Coxeter automorphism of $L(\mathfrak{g}, C)$ if $\mathfrak{g}^0$ is Abelian and $C$ is of minimal order. The number $r=\mbox{ dim}(\mathfrak{g}^0)$ is called the rank of $L(\mathfrak{g}, C)$. The order $h$ of $C$ is called the Coxeter number of $L(\mathfrak{g}, C)$. We will be using Coxeter automorphisms to construct the Kac-Moody algebras needed in this work.

Let $p$ is a permutation of the simple roots that preserves the Dynkin diagram of $\mathfrak{g}$. Every such $p$ induces an outer automorphism $P$ of $\mathfrak{g}$.
Assume that a Coxeter automorphism $C$ of a Kac-Moody algebra $L(\mathfrak{g}, C)$ is of the form
\begin{equation}
C= \tilde{C} \circ P,
\end{equation}
where $\tilde{C}$ is an inner automorphism induced by a Weyl group element group element $\tilde{c}$.
\footnote{
Not every Coxeter automorphism can be realized in this way. For example, Coxeter automorphisms of the form $C(X) = c F(X) c^{-1}$, where $c$ is a diagonal matrix and $F$ is some properly chosen outer automorphism, can never be constructed from a Weyl group element.
}
Every such $C$ is induced by a linear mapping $c = \tilde{c} \circ p$ acting in the root space $\mathfrak{g}$.
By analogy with simple Lie algebras, the eigenvalues of $c$ are called exponents of the Kac-Moody algebra $L(\mathfrak{g}, C)$.

A basis of $L(\mathfrak{g}, C)$  can be constructed as follows:
Each element X of $L(\mathfrak{g}, C)$ is of the form
\begin{equation}
X=\sum_{k= n}^{m} X^{(k)} \lambda^{k}, \quad n,m \in \mathbb{Z},
\end{equation}
where $X^{(k)} \in \mathfrak{g}^{(k \mbox{ mod } h)}$.
Each of the subspaces $\mathfrak{g}^{(k)}$ has a basis given by:
\begin{equation}
 \mathcal{E}_{\alpha}^{(k)} =  \sum_{s=0}^{h-1} \omega^{-s k} C^{s} ( E_{\alpha}), \qquad  \mathcal{H}_j^{(k)} = \sum_{s=0}^{h-1} \omega^{-s k} C^{s} ( H_{j}).
\label{basis1}
\end{equation}
Note that $\mathcal{H}_j^{(k)}$ is non-vanishing only if $k$ is an exponent. This means that the number of elements in $\mathfrak{g}^{(k)}$ is $r+1$ if $k$ is an exponent and $r$ otherwise, where $r$ is the rank of $L(\mathfrak{g}, C)$.
The roots $\alpha$ are chosen as follows:
$c$ splits the root system of $\mathfrak{g}$ into $r$ non-intersecting orbits. From each orbit we select only one root $\alpha$.

This work is mainly concerned with Kac-Moody algebras of type $D_4$. Note that, since there are two types of outer automorphisms of $D_4$, there are three Kac-Moody algebras - $D_4^{(1)}$,$D_4^{(2)}$ and $D_4^{(3)}$ (here the upper index denotes the height of the algebra).
The Coxeter numbers and the exponents of those algebras are given in Table \ref{tab:1} (the values are taken from \cite{DriSok}).
The table also contains the Coxeter automorphisms we use. The construction of each of those automorphisms is given in the corresponding section.
\begin{table}
  \centering
\begin{tabular}{|c|c|c|c|c|}
  \hline
   Algebra & Coxeter automorphism & Coxeter number &Exponents &Rank  \\
  \hline
$  D_4^{(1)} $ & $C_1 = S_{\alpha_2}  S_{\alpha_1}  S_{\alpha_3}  S_{\alpha_4}$ & $6$ & $1,3,5,3 $ & $4$\\
$  D_4^{(2)} $ & $C_2 = S_{\alpha_1}  S_{\alpha_3}  S_{\alpha_2} R $ & $8$ & $ 1,3,5,7 $ & $3$\\
$  D_4^{(3)} $ & $C_3 = S_{\alpha_2}  S_{\alpha_1}  T $ & $12$ & $ 1,5,7,11 $ & $2$ \\
     \hline
\end{tabular}
\caption{ \textit{ A realization of the Coxeter automorphisms and Coxeter numbers for $D_4^{(1)}, D_4^{(2)}$ and  $D_4^{(3)}$. Here $S_{\alpha_i}$
denotes reflection with respect to the simple root $\alpha_i$, $R$ is the second order outer automorphism that exchanges $\alpha_3$ and $\alpha_4$,
and $T$ is the third order outer automorphism (a triality transformation) that sends $\alpha_1 \rightarrow \alpha_3 \rightarrow \alpha_4$.} }
\label{tab:1}
\end{table}
The explicit form of the basis for each Kac-Moody algebra of type $D_4$ is given below.
%


\subsection{The simple Lie algebra $D_4$}
We will review the most important properties of the simple Lie algebra $D_4 \equiv \mathfrak{so}(8)$.
Let $e_i$ be a standard basis in the root space of $D_4$ (from now on, if we don't specify otherwise, we will always assume that this is the basis of the root space). For the root system of $D_4$ we have $\Delta = \Delta_+ \cup \Delta_-$, where
\begin{equation}
\begin{aligned}
&\Delta_+  = \left\{   e_i \pm e_j: i,j =1...4, \, i<j, \right\}, \qquad \Delta_-  =  \left\{   -(e_i \pm e_j): i,j =1...4, \, i<j. \right\}
\end{aligned}
\end{equation}

\begin{figure}
\begin{center}
\includegraphics[width=0.22\textwidth]{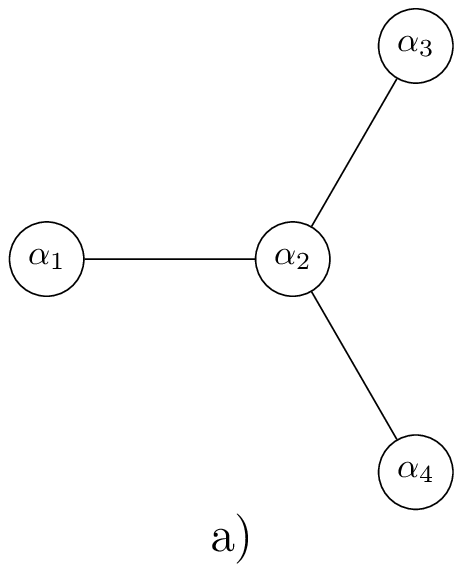} \qquad \includegraphics[width=0.25\textwidth]{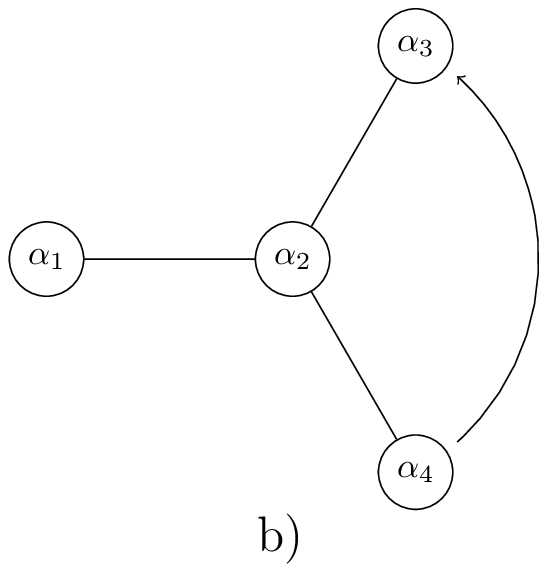} \qquad
\includegraphics[width=0.25\textwidth]{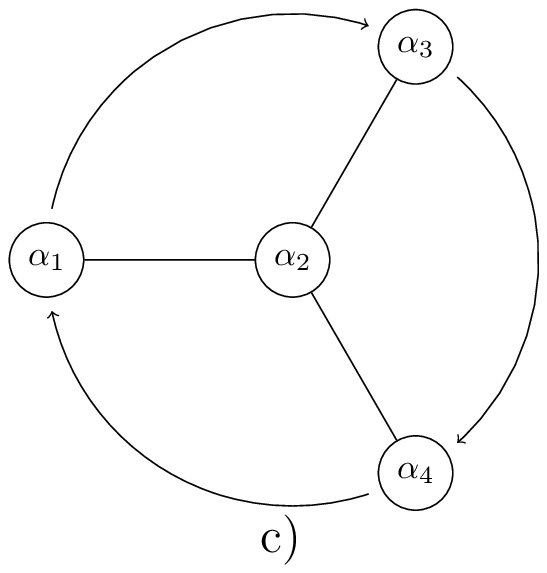}
\end{center}
\caption{\it a) Dynkin diagram of the simple Lie algebra $D_4$; a) Dynkin diagram of the simple Lie algebra $D_4$ with the mirror automorphism $R$;
a) Dynkin diagram of the simple Lie algebra $D_4$ with the third order outer automorphism $T$.}
\label{DDD4}
\end{figure}

The simple roots are given by
\begin{equation}
\begin{aligned}
&\alpha_1 = e_1 - e_2, \quad \alpha_2 = e_2 - e_3, \qquad \alpha_3 = e_3 - e_4, \quad \alpha_4 = e_3 + e_4.
\end{aligned}
\end{equation}
With each vertex of the Dynkin diagram of $D_4$ we can associate a simple root (Fig \ref{DDD4}).

$D_4$ is usually represented by a $8\times 8$ antisymmetric matrices.
In this representation the Cartan subalgebra is not diagonal, so we will use a representation for which every $X \in D_4$ satisfies
\begin{equation}
SX+(SX)^T=0,
\end{equation}
where the matrix $S$ is given by
\begin{equation}
S=
\begin{pmatrix}
0 & 0 & 0 & 0 & 0 & 0 & 0 & 1 \\
0 & 0 & 0 & 0 & 0 & 0 & -1 & 0 \\
0 & 0 & 0 & 0 & 0 & 1 & 0 & 0  \\
0 & 0 & 0 & 0 & -1 & 0 & 0 & 0 \\
0 & 0 & 0 & -1 & 0 & 0 & 0 & 0 \\
0 & 0 & 1 & 0 & 0 & 0 & 0 & 0 \\
0 & -1 & 0 & 0 & 0 & 0 & 0 & 0 \\
1 & 0 & 0 & 0 & 0 & 0 & 0 & 0
\end{pmatrix}.\label{S}
\end{equation}
This way the Cartan subalgebra is given by diagonal matrices.
The Cartan-Weyl generators of $D_4$ are given by
\begin{equation}
\begin{aligned}
H_i &= e_{ii} - e_{9-i,9-i},  \quad  i =1,\dots, 4, &\; E_{\alpha_j} &= e_{j, j+1}+e_{8-j,9-j}, \; j = 1,2, 3, &\; E_{\alpha_4} &= e_{3,5}+ e_{4,6}, \\
E_{e_i-e_j}&= e_{i,j} -(-1)^{i+j} e_{9-j,9-i}, &\; E_{e_i+e_j} &= e_{i,9-j} -(-1)^{i+j} e_{j,9-i},, &\;  E_{-\alpha_j}&=(E_{\alpha_j})^T,
\end{aligned}
\end{equation}
where $  1 \leq i < j \leq 4$. By $e_{ij}$ we denote a matrix that has a one at the $i-th$ row and $j-th$ column and is zero everywhere else.
The exponents of $D_4$ are $1,3,3,5$ and its Coxeter number is $6$.

\subsection{The Coxeter automorphisms of $D_4^{(s)}$}

In what follows by  $S_{\alpha_i}$ we will denote the Weyl reflection with respect to the simple root $\alpha_i$.
It is well known that each element of the Weyl group naturally induces an inner automorphism.
\begin{equation}
S_{\alpha_i}(E_\beta) = s_i E_\beta s_i^{-1} =n_{\alpha_i,\beta} E_{S_{\alpha_i}(\beta)}, \qquad S_{\alpha_i}(H_\beta)= H_{S_{\alpha_i}(\beta)}
\end{equation}
where $n_{\alpha_i,\beta} =\pm 1$ and the matrices $s_i$ are easy to find.

The Coxeter automorphism for $D_4^{(1)}$ coincides with the Coxeter automorphism for $D_4$ and is given by
\begin{equation}
C_1 = S_{\alpha_2} S_{\alpha_1} S_{\alpha_3} S_{\alpha_4}.
\end{equation}
In root space it has the form
\begin{equation}
C_1=\begin{pmatrix}
0 & 1 & 0 & 0 \\
0 & 0 & -1 & 0 \\
1 & 0 & 0 & 0 \\
0 & 0 & 0 & -1
\end{pmatrix}.
\label{coxeter1_RS}
\end{equation}
$C_1$ splits the root system of $D_4$ into 4 orbits, each containing 6 elements:
\begin{equation*}
\begin{array}{cccccccccccc}
\mathcal{O}_1 \colon & e_1 - e_2  &\to& -(e_1 - e_3) &\to& -(e_2 + e_3) &\to& -(e_1 - e_2) &\to& e_1 - e_3 &\to& e_2 + e_3, \\
\mathcal{O}_2 \colon & e_2 - e_3  &\to&  \; e_1 + e_2   &\to& e_1 + e_3 &\to& -(e_2 - e_3) &\to& -(e_1 + e_2) &\to&  -(e_1 + e_2), \\
\mathcal{O}_3 \colon & e_3 - e_4  &\to& -(e_2 - e_4) &\to& -(e_1 + e_4) &\to& -(e_3-e_4) &\to& e_2 - e_4   	&\to& e_1 + e_4, \\
\mathcal{O}_4 \colon & e_3 + e_4 &\to& -(e_2 + e_4) &\to& - (e_1 - e_4) &\to& -(e_3 + e_4) &\to& e_2 + e_4 &\to& e_1 - e_4.
\end{array}
\end{equation*}
When building the basis \eqref{basis1} we will average only the Weyl generators corresponding to the simple roots $\alpha_1 = e_1 - e_2$, $\alpha_2 = e_2 - e_3$, $\alpha_3 = e_3 - e_4$, $\alpha_4 = e_3 + e_4$. In the algebra $C_1$ is realized as an inner automorphism , i.e. a similarity transformation
\begin{equation}
C_1(X) = c_1 X c_1^{-1} , \qquad X \in D_4^{(1)}, \qquad c_1 = \begin{pmatrix}
0 & -1 & 0 & 0 & 0 & 0 & 0 & 0 \\
0 & 0 & 0 & 0 & 0 & -1 & 0 & 0 \\
1 & 0 & 0 & 0 & 0 & 0 & 0 & 0  \\
0 & 0 & 0 & 0 &-1 & 0 & 0 & 0 \\
0 & 0 & 0 &- 1 & 0 & 0 & 0 & 0 \\
0 & 0 & 0 & 0 & 0 & 0 & 0 & 1 \\
0 & 0 & 1 & 0 & 0 & 0 & 0 & 0 \\
0 & 0 & 0 & 0 & 0 & 0 & 1 & 0
\end{pmatrix}. \label{coxeter1}
\end{equation}
It is easy to check that $C_1^6=\openone$.

We will also use the analog of the Cartan-Weyl basis in each of the subspace $\mathfrak{g}^{(k)}$ by taking
weighted average of $E_{\alpha_i}$ over the action of the Coxeter automorphism:
\begin{equation}\label{eq:EakC1}\begin{aligned}
 \mathcal{E}_i^{(k)} &= \sum_{s=0}^{5} \omega_1^{-sk} C_1^s(E_{\alpha_i}), \qquad  \mathcal{H}_i^{(k)} &= \sum_{s=0}^{5} \omega_1^{-sk} C_1^s(H_{e_i}).
\end{aligned}\end{equation}
One can check that both $ \mathcal{E}_i^{(k)}$ and $ \mathcal{H}_i^{(k)}$ must belong to $\mathfrak{g}^{(k)}$; indeed, it is
easy to see that $C_1( \mathcal{E}_i^{(k)})=\omega_1^k  \mathcal{E}_i^{(k)}$ and $C_1( \mathcal{H}_i^{(k)})=\omega_1^k  \mathcal{H}_i^{(k)} $.
Another important remark is that $ \mathcal{H}_i^{(k)}$ is not vanishing if and only if $k$ is an exponent of $D_4^{(1)}$. In addition
for $k=1$ and $k=5$ it is enough to consider only $ \mathcal{H}_1^{(k)}$; considering $ \mathcal{H}_i^{(k)}$ for $i=2,3,4$ we find that they
are proportional to $ \mathcal{H}_1^{(k)}$. The exception here is only for the case $k=3$; then we have two linearly independent
Cartan elements: $ \mathcal{H}_1^{(k)}$ and $ \mathcal{H}_4^{(k)}$. Skipping the details we list the results:
\begin{equation}\label{eq:EakHk1}\begin{aligned}
\mathcal{E}_1^{(k)} &=  E_{e_1-e_2} - \omega_1^{-p} E_{-e_1+e_3}-\omega_1^{-2p} E_{-e_2-e_3}-\omega_1^{-3p} E_{-e_1+e_2}  +\omega_1^{-4p} E_{e_1-e_3} +\omega_1^{-5p} E_{e_2+e_3}, \\
\mathcal{E}_2^{(k)} &= E_{e_2-e_3} - \omega_1^{-p} E_{e_1+e_2}+\omega_1^{-2p} E_{e_1+e_3}-\omega_1^{-3p} E_{-e_2+e_3}  +\omega_1^{-4p} E_{-e_1-e_2} -\omega_1^{-5p} E_{-e_1-e_3}, \\
\mathcal{E}_3^{(k)} &=  E_{e_3-e_4} - \omega_1^{-p} E_{-e_2+e_4}+\omega_1^{-2p} E_{-e_1-e_4}+\omega_1^{-3p} E_{-e_3+e_4}  -\omega_1^{-4p} E_{e_2-e_4} +\omega_1^{-5p} E_{e_1+e_4}, \\
\mathcal{E}_4^{(k)} &= E_{e_3+e_4} - \omega_1^{-p} E_{-e_2-e_4}+\omega_1^{-2p} E_{-e_1+e_4}+\omega_1^{-3p} E_{-e_3-e_4}  -\omega_1^{-4p} E_{e_2+e_4} +\omega_1^{-5p} E_{e_1-e_4},  \\
 \mathcal{H}_1^{(s)} &=  2(H_{e_1} +\omega_1^{s} H_{e_2} +\omega_1^{-s} H_{e_3}), \qquad  \mathcal{H}_4^{(3)} = 6H_{e_4}, \qquad s=1, 3, 5.
\end{aligned}\end{equation}
These results allow us also to calculate the commutation relations between the basis in (\ref{eq:EakHk1}). In particular
\begin{equation}\label{eq:adHk1}\begin{split}
 {} [ \mathcal{H}_1^{(s)} ,\mathcal{E}_i^{(k)}] =  \alpha_i(\mathcal{H}_1^{(s)}) \mathcal{E}_i^{(k+s)}, \qquad [ \mathcal{H}_4^{(3)} ,\mathcal{E}_i^{(k)}] =  \alpha_i(\mathcal{H}_4^{(3)}) \mathcal{E}_i^{(k+3)}.
\end{split}\end{equation}
where
\begin{equation}\label{eq:alfaHk}
\begin{aligned}
&\alpha_1(\mathcal{H}_1^{(s)}) =2(1 - \omega_1^s ) , \quad
\alpha_2(\mathcal{H}_1^{(s)}) = 2(\omega_1^s - \omega_1^{-s}) ,  \quad
\alpha_3(\mathcal{H}_1^{(s)}) =2 \omega_1^{-s} , \quad
\alpha_4(\mathcal{H}_1^{(s)}) =2 \omega_1^{-s} ,  \\
&\alpha_1(\mathcal{H}_4^{(3)}) = 0 , \quad
\alpha_2(\mathcal{H}_4^{(3)}) = 0 , \quad
\alpha_3(\mathcal{H}_4^{(3)}) = -6 ,\quad
\alpha_4(\mathcal{H}_4^{(3)}) =  6.
\end{aligned}
\end{equation}

The Coxeter automorphism for $D_4^{(2)}$ is given by
\begin{equation}
C_2 = S_{\alpha_1}  S_{\alpha_3}  S_{\alpha_2} R,
\end{equation}
where $R$ is the outer automorphism that exchanges $\alpha_3$ and $\alpha_4$.
In the root space of $G$ we have
\begin{equation}\label{eq:C2}
C_2=\begin{pmatrix}
0 & 0 & 1 & 0 \\
1 & 0 & 0 & 0 \\
0 & 0 & 0 & -1 \\
0 & 1 & 0 & 0
\end{pmatrix}.
\end{equation}
From (\ref{eq:C2}) it is easy to check that  $C_2^8=\openone $ and its eigenvalues are $\omega_2^1,\omega_2^3,\omega_2^5,\omega_2^7$,
where $\omega_2=\exp(2\pi i/8)$; compare with the Coxeter number and the exponents for $D_4^{(2)}$ in Table \ref{tab:1}.
$C_2$ splits the root system of $D_4$ into 3 orbits, each containing 8 elements:
\begin{equation*}
\begin{array}{cccccccccccccc}
\mathcal{O}_1  \colon & e_1 - e_2 &\to& e_2 - e_4  &\to& e_3 + e_4 &\to& e_1 - e_3 &\to& -(e_1 - e_2) &\to& -(e_2 - e_4)& \to  \\
 & -(e_3 + e_4) &\to& -(e_1 - e_3),& & & & & & & & &  \\[3pt]
\mathcal{O}_2  \colon & e_2 - e_3 &\to& -(e_1 - e_4) &\to& -(e_2 + e_3) &\to& -(e_1 + e_4) &\to& -(e_2 - e_3) &\to& e_1 - e_4& \to\\
& e_2 + e_3 &\to& e_1 + e_4, & & & & & & & & & \\[3pt]
\mathcal{O}_3  \colon & e_3 - e_4 &\to& e_1 + e_3 &\to& e_1 + e_2 &\to& e_2 + e_4 &\to& -(e_3 - e_4) &\to& -(e_1 + e_3) &\to \\
& -(e_1 + e_2) &\to& -(e_2 + e_4) & & & & & & & & & .
\end{array}
\end{equation*}
The roots chosen in \eqref{basis1} are $\alpha_1 = e_1 - e_2$, $\alpha_2 = e_2 - e_3$ , $\alpha_3 = e_3 - e_4$.

 In the algebra $R$ is realized as a similarity transformation with a matrix $r$, given by
\begin{equation}
r =
\begin{pmatrix}
1 & 0 & 0 & 0 & 0 & 0 & 0 & 0 \\
0 & 1 & 0 & 0 & 0 & 0 & 0 & 0 \\
0 & 0 & 1 & 0 & 0 & 0 & 0 & 0  \\
0 & 0 & 0 & 0 & 1 & 0 & 0 & 0 \\
0 & 0 & 0 & 1 & 0 & 0 & 0 & 0 \\
0 & 0 & 0 & 0 & 0 & 1 & 0 & 0 \\
0 & 0 & 0 & 0 & 0 & 0 & 1 & 0 \\
0 & 0 & 0 & 0 & 0 & 0 & 0 & 1
\end{pmatrix}.
\end{equation}
The Coxeter automorphism of $D_4^{(2)}$ is realized as
\begin{equation}
C_2(X) = c_2 X c_2^{-1} , \qquad X \in D_4^{(2)}, \qquad c_2 =
\begin{pmatrix}
0 & 0 & 1 & 0 & 0 & 0 & 0 & 0 \\
1 & 0 & 0 & 0 & 0 & 0 & 0 & 0 \\
0 & 0 & 0 & 0 & 1 & 0 & 0 & 0  \\
0 & 1 & 0 & 0 & 0 & 0 & 0 & 0 \\
0 & 0 & 0 & 0 & 0 & 0 & 1 & 0 \\
0 & 0 & 0 & -1 & 0 & 0 & 0 & 0 \\
0 & 0 & 0 & 0 & 0 & 0 & 0 & -1 \\
0 & 0 & 0 & 0 & 0 & 1 & 0 & 0
\end{pmatrix}.
\end{equation}
The analog of the Cartan-Weyl basis in  the subspaces $\mathfrak{g}^{(k)}$ is given by
\begin{equation}\label{eq:EakC2}\begin{aligned}
 \mathcal{E}_i^{(k)} &= \sum_{s=0}^{7} \omega_2^{-sk} C_2^s(E_{\alpha_i}), \qquad  \mathcal{H}_i^{(k)} &= \sum_{s=0}^{5} \omega_2^{-sk} C_2^s(H_{e_i}).
\end{aligned}\end{equation}
where we take only the roots $\alpha_1$, $\alpha_2$ and $\alpha_3$; each of them specifies a different orbit of $C_2$.
Obviously that $C_2( \mathcal{E}_i^{(k)})=\omega_2^k  \mathcal{E}_i^{(k)}$ and $C_2( \mathcal{H}_i^{(k)})=\omega_2^k  \mathcal{H}_i^{(k)} $.
Besides $ \mathcal{H}_i^{(k)}$ is not vanishing if and only if $k$ is an exponent of $D_4^{(1)}$.
It is enough to consider only $ \mathcal{H}_1^{(k)}$; the Cartan elements $ \mathcal{H}_i^{(k)}$ for $i=2,3,4$
are proportional to $ \mathcal{H}_1^{(k)}$.  Skipping the details we list the results:
\begin{equation}\label{eq:EakHk2}\begin{aligned}
\mathcal{E}_1^{(k)} &=  E_{e_1-e_2} + \omega_2^{-p} E_{e_2-e_4}+\omega_2^{-2p} E_{e_3+e_4}-\omega_2^{-3p} E_{e_1-e_3}  -\omega_2^{-4p} E_{-e_1+e_2} -\omega_2^{-5p} E_{-e_2+e_4} \\
 &- \omega_2^{-6p} E_{-e_3-e_4} +\omega_2^{-7p} E_{-e_1+e_3}, \\
\mathcal{E}_2^{(k)} &= E_{e_2-e_3} + \omega_2^{-p} E_{-e_1+e_4}+\omega_2^{-2p} E_{-e_2-e_3}+\omega_2^{-3p} E_{-e_1-e_4}  -\omega_2^{-4p} E_{-e_2+e_3} -\omega_2^{-5p} E_{e_1-e_4} \\
 &- \omega_2^{-6p} E_{e_2+e_3} -\omega_2^{-7p} E_{e_1+e_4}, \\
  \mathcal{E}_3^{(k)} &= E_{e_3-e_4} + \omega_2^{-p} E_{e_1+e_3}+\omega_2^{-2p} E_{e_1+e_2}+\omega_2^{-3p} E_{e_2+e_4}  -\omega_2^{-4p} E_{-e_3+e_4} -\omega_2^{-5p} E_{-e_1-e_3} \\
 &- \omega_2^{-6p} E_{-e_1-e_2} -\omega_2^{-7p} E_{-e_2-e_4},  \\
\mathcal{H}_1^{(p)} &=  2(H_{e_1} +\omega_2^{-p} H_{e_2} -\omega_2^{-3p} H_{e_3} +\omega_2^{-2p} H_{e_4}).
\end{aligned}\end{equation}
These results allow us also to calculate the commutation relations between the basis in (\ref{eq:EakHk2}). In particular
\begin{equation}\label{eq:adHk2}\begin{split}
 {} [ \mathcal{H}_1^{(s)} ,\mathcal{E}_i^{(k)}] =  \alpha_i(\mathcal{H}_1^{(s)}) \mathcal{E}_i^{(k+s)}, \qquad
\end{split}\end{equation}
where
\begin{equation}\label{eq:alfaH2}\begin{split}
\alpha_1(\mathcal{H}_1^{(s)}) &= 2(1-\omega_2^{-s}) , \qquad \alpha_2(\mathcal{H}_1^{(s)}) =  2(\omega_2^{-s}-\omega_2^{s}) , \\
\alpha_3(\mathcal{H}_1^{(s)}) &=  2(\omega_2^{s}-\omega_2^{-2s}) , \qquad
\alpha_4(\mathcal{H}_1^{(s)}) =  2(\omega_2^{s}+\omega_2^{-2s}) ,
\end{split}\end{equation}

The Coxeter element for $D_4^{(3)}$ is given as
\begin{equation}
C_3 = S_{\alpha_2}  S_{\alpha_1}  T,
\end{equation}
where $T$ (known in some literature as triality transformation) is the third order outer automorphism for which
\begin{equation}
T: \alpha_1 \mapsto \alpha_3 \mapsto \alpha_4
\end{equation}
and is stationary on $\alpha_2$. In root space $T$ can be realized as a matrix
\begin{equation}
T=\frac{1}{2} \begin{pmatrix}
1 & 1 & 1 & 1 \\
1 & 1 & -1 & -1 \\
1 & -1 & 1 & -1 \\
-1 & 1 & 1 & -1
\end{pmatrix}.
\end{equation}

 Unfortunately, it seems that there is no similarity transformation that realizes $T$ in the algebra. By knowing the action on the Weyl generators $E_{\alpha_i}$ one can construct the action over the whole algebra using the fact that
\begin{equation}
T \left( \big[ E_{\alpha_i} , E_{\alpha_j} \big] \right) =  \big[ T(E_{\alpha_i}) , T(E_{\alpha_j}) \big].
\end{equation}
This leads to
\begin{equation}\label{eq:V23}
\begin{aligned}
T &\colon E_{\alpha_1} \to  E_{\alpha_3} \to E_{\alpha_4}, \\
T &\colon E_{\alpha_1+\alpha_2} \to - E_{\alpha_2+\alpha_3} \to -E_{\alpha_2+\alpha_4}, \\
T &\colon E_{\alpha_1+\alpha_2+\alpha_3} \to - E_{\alpha_2+\alpha_3+\alpha_4} \to E_{\alpha_1+\alpha_2+\alpha_4},
\end{aligned}
\end{equation}
with stationary elements $E_{\alpha_2}$, $E_{\alpha_1 + \alpha_2 + \alpha_3 + \alpha_4}$ and $E_{\alpha_1 + 2\alpha_2 + \alpha_3 + \alpha_4}$
The action on the negative roots is obtained by transposing the above.
Thus we obtain the following realization of $C_3$ in the root space of $D_4$.
\begin{equation}\label{eq:C3}
C_3=\frac{1}{2}
\begin{pmatrix}
1 & 1 & -1 & -1 \\
1 & -1 & 1 & -1 \\
1 & 1 & 1 & 1 \\
-1 & 1 & 1 & -1
\end{pmatrix}.
\end{equation}
From (\ref{eq:C3}) it is easy to check that  $C_3^{12}=\openone $ and its eigenvalues are $\omega_3^1,\omega_3^5,\omega_3^7,\omega_3^{11}$,
where $\omega_3=\exp(2\pi i/12)$; compare with the Coxeter number and the exponents for $D_4^{(3)}$ in Table \ref{tab:1}.

Therefore $C_3$ splits the roots  of $D_4$ into 2 orbits, each containing  12 elements:
\begin{equation*}
\begin{array}{ccccccccccccc}
\mathcal{O}_1  \colon & e_2 - e_3 &\to & e_1 - e_2 &\to& e_2 - e_4 &\to& e_1 + e_4 &\to& e_3 - e_4 &\to& e_2 + e_4 &\to  \\
& -(e_2 - e_3) &\to & -(e_1 - e_2) &\to& -(e_2 - e_4) &\to& -(e_1 + e_4) &\to& -( e_3 - e_4) &\to& -(e_2 + e_4) & \\[3pt]
\mathcal{O}_2  \colon & e_3 + e_4 &\to& -(e_1 - e_3) &\to& -(e_1 - e_4) &\to& -(e _1 + e_2) &\to& -(e_1 + e_3) &\to& -(e_2 + e_3) &\to  \\
& -(e_3 + e_4) &\to& e_1 - e_3 &\to& e_1 - e_4 &\to& e_1 + e_2 &\to& e_1 + e_3 &\to& e_2 + e_3.&
\end{array}
\end{equation*}
In \eqref{basis1} we choose the roots $\alpha_1 = e_1-e_2$ and $\alpha_4 = e_3+e_4$.

The Coxeter automorphism of $D_4^{(3)}$ is then realized as
\begin{equation}
C_3(X) = S_2 S_1 T(X) S_1 S_2.
\end{equation}

The  Cartan-Weyl basis in  the subspaces $\mathfrak{g}^{(k)}$ takes the form
\begin{equation}\label{eq:EakC3}\begin{aligned}
 \mathcal{E}_i^{(k)} &= \sum_{s=0}^{11} \omega_3^{-sk} C_3^s(E_{\alpha_i}), \qquad  \mathcal{H}_i^{(k)} &= \sum_{s=0}^{11} \omega_3^{-sk} C_3^s(H_{e_i}).
\end{aligned}\end{equation}
where we take only the roots $\alpha_2$ and $\alpha_4$; each of them specifies a different orbit of $C_3$.
Obviously that $C_3( \mathcal{E}_i^{(k)})=\omega_3^k  \mathcal{E}_i^{(k)}$ and $C_3( \mathcal{H}_i^{(k)})=\omega_3^k  \mathcal{H}_i^{(k)} $.
Besides $ \mathcal{H}_i^{(k)}$ is not vanishing if and only if $k$ is an exponent of $D_4^{(3)}$.
It is enough to consider only $ \mathcal{H}_1^{(k)}$; the Cartan elements $ \mathcal{H}_i^{(k)}$ for $i=2,3,4$
are proportional to $ \mathcal{H}_1^{(k)}$.  Skipping the details we list the results:
\begin{equation}\label{eq:EakHk3}\begin{aligned}
&\mathcal{E}_2^{(k)} =
E_{e_2-e_3} + \omega_3^{-p} E_{e_1-e_2}+\omega_3^{-2p} E_{e_2-e_4}+\omega_3^{-3p} E_{e_1+e_4}  +\omega_3^{-4p} E_{e_3-e_4} +\omega_3^{-5p} E_{e_2+e_4} \\
&-\omega_3^{-6p} E_{-e_2+e_3} - \omega_3^{-7p} E_{-e_1+e_2} - \omega_3^{-8p} E_{-e_2+e_4}-\omega_3^{-9p} E_{-e_1-e_4}  -\omega_3^{-10p} E_{-e_3+e_4} -\omega_3^{-11p} E_{-e_2-e_4}, \\
&\mathcal{E}_4^{(k)} =
E_{e_3+e_4} + \omega_3^{-p} E_{-e_1+e_3}-\omega_3^{-2p} E_{-e_1+e_4}-\omega_3^{-3p} E_{-e_1-e_2}  -\omega_3^{-4p} E_{-e_1-e_3} +\omega_3^{-5p} E_{-e_2-e_3} \\
& -\omega_3^{-6p} E_{-e_3-e_4} - \omega_3^{-7p} E_{e_1-e_3} + \omega_3^{-8p} E_{e_1-e_4}+\omega_3^{-9p} E_{e_1+e_2}  -\omega_3^{-10p} E_{e_1+e_3} -\omega_3^{-11p} E_{e_2+e_3}, \\
 &\mathcal{H}_1^{(1)} = \sqrt{3} \left(\begin{array}{c} \sqrt{3} +1 \\ 1-i \\  (\sqrt{3}+1) i \\ -1 -i \end{array}\right), \quad
 \mathcal{H}_1^{(5)} = \sqrt{3} \left(\begin{array}{c} \sqrt{3}-1 \\ -1+i \\  -(\sqrt{3}-1) i \\ 1 +i \end{array}\right), \quad
 \mathcal{H}_1^{(7)} = \mathcal{H}_1^{(5),*}, \quad
  \mathcal{H}_1^{(11)} = \mathcal{H}_1^{(1),*}.
\end{aligned}\end{equation}
These results allow us also to calculate the commutation relations between the basis in (\ref{eq:EakHk3}). In particular
\begin{equation}\label{eq:adHk3}\begin{split}
 {} [ \mathcal{H}_1^{(s)} ,\mathcal{E}_i^{(k)}] =  \alpha_i(\mathcal{H}_1^{(k)}) \mathcal{E}_i^{(k+s)}, \qquad
\end{split}\end{equation}
where
\begin{equation}\label{eq:alfaH3}\begin{split}
\alpha_2(\mathcal{H}_1^{(s)}) =  \begin{cases} \sqrt{3}(1 + i\sqrt{3}) & \mbox{for}, \; s=1, \\ \sqrt{3}(-1 + i\sqrt{3}) & \mbox{for} \; s=5,
\\  \sqrt{3}(-1 - i\sqrt{3}) & \mbox{for} \; s=7, \\  \sqrt{3}(1 - i\sqrt{3}) & \mbox{for} \; s=11,\end{cases}
, \qquad \alpha_4(\mathcal{H}_1^{(s)}) = \begin{cases} \sqrt{3}(-1 - i(2+\sqrt{3})) & \mbox{for} \; s=1, \\ \sqrt{3}(1 + i(2-\sqrt{3})) & \mbox{for} \; s=5,
\\  \sqrt{3}(1 - i(2-\sqrt{3})) & \mbox{for} \; s=7, \\  \sqrt{3}(-1 + i(2+\sqrt{3})) & \mbox{for} \; s=11.\end{cases}
\end{split}\end{equation}

\begin{remark}\label{rem:1}
In order to simplify the usage of indices,  we have used the same letters and types of indices to denote the Cartan-Weyl basis in $\mathfrak{g}^{(k)}$
for each of the Kac-Moody algebras $D_4^{(s)}$. Of course it will be clear from the context which of them we have in mind.
\end{remark}




\section{Lax representations and recursion operators}

\subsection{Lax representations and reductions}
Let us consider a generic Lax pair,
\begin{equation}\label{LaxPair}
\begin{aligned}
L \psi &\equiv (i \partial_x + U(x,t,\lambda))\psi(x,t,\lambda)=0, &\quad
M \psi &\equiv (i \partial_t + V(x,t,\lambda))\psi(x,t,\lambda)=0,  \\
U(x,t,\lambda) &=Q(x,t) - \lambda J, &\quad
V(x,t,\lambda) &= \sum_{k=0}^{n-1} \lambda^k V_{k}(x,t) - \lambda^{n}K,
\end{aligned}
\end{equation}
whose potentials $U(x,t,\lambda)$  and $V(x,t,\lambda)$ are elements of the Kac-Moody algebra $D_4^{(s)}$;
in fact most of the results in this subsection will be  valid for any Kac-Moody algebra.
This means that
\begin{equation}
\label{potential}
Q(x,t) \in \mathfrak{g}^{(0)}, \quad V_{k}(x,t) \in \mathfrak{g}^{(k)}, \quad K \in \mathfrak{g}^{(n)}\cap \mathfrak{h}, \quad
J\in \mathfrak{g}^{(1)}\cap \mathfrak{h}.
\end{equation}
Such choice for the potentials of the Lax pair means that they involve $\mathbb{Z}_h$ as their reduction group   \cite{Mikhailov}:
\begin{equation}
C\left( U(x,t,\lambda) \right) = U(x,t,\omega\lambda),\quad C\left( V(x,t,\lambda) \right) = V(x,t,\omega\lambda).
\label{Red1}
\end{equation}
with $\omega =e^{\frac{2\pi i}{h}}$ is the Coxeter number. We also assume that $J$ and $K$ are constant elements of the Cartan
subalgebra.

We request that the operators $L$ and $M$ commute identically with respect to $\lambda$. In particular, since  $[J,K]=0$  this means that
the leading power $n$ in $V(x,t,\lambda)$ must be of the form $n= n_0h +n_1$, where $n_1$ must be an exponent of $D_4^{(s)}$.
To simplify the notation we will often omit writing the explicit dependence on $x$ and $t$.
This implies the following recursion relations
\begin{equation}
\begin{aligned}
&\lambda^{n+1}: & \big[ J, K \big]&=0, \\
&\lambda^{n}:    & \big[ J,V_{n-1} \big] + \big[ Q, K \big]&=0, \\
&\lambda^{n-1}:    &  i{\partial_x V_{n-1}}+ \big[Q ,V_{n-1} \big] &= \big[ J,V_{n-2} \big] , \\
&\lambda^{s}:     &  i{\partial_x V^{(s)}}+ \big[Q ,V_{s} \big]&= \big[ J,V_{s-1} \big] , \\
&\lambda^{0}:    &  - i{\partial_t Q}+  i{\partial_x V_{0}}+  \big[ Q(x,t),V_{0} \big]&=0. \\
\end{aligned} \label{Recurrence1}
\end{equation}
Now we view eq. (\ref{Recurrence1}) as a set of recurrence relations and aim to resolve them and express all $V_s(x,t)$ in terms of
$Q(x,t)$. Doing this we have to take into account that the operator $\ad_J X \equiv [ H, X] $ has a kernel. Therefore we need
to split each $V_{s}$ into a sum of diagonal and off-diagonal parts. Remembering the results of the previous section we set $s=s_0h+s_1$
and consider two cases:
\begin{equation}\label{eq:Vsrec}\begin{split}
V_{s}(x,t) = \begin{cases} V_{s}^{\rm f}(x,t) & \mbox{if}\; s_1 \; \mbox{is not an exponent} \\
 V_{s}^{\rm f}(x,t) + w_{s}(x,t)\mathcal{H}_1^{(s_1)} & \mbox{if}\; s_1 \; \mbox{is  an exponent}
\end{cases}, \qquad V_s^{\rm f}(x,t) = \sum_{p=1}^{r} V_{s,p}(x,t) \mathcal{E}_p^{(s)} .
\end{split}\end{equation}

\begin{remark}\label{rem:2}
 Note that using proper gauge transformation we can always transform away the diagonal part of $V_{n-1}$; so
 using the commutation relations (\ref{eq:adHk1}), (\ref{eq:adHk2}) or (\ref{eq:adHk3}) From the second of the equations  (\ref{Recurrence1}) we obtain:
 \begin{equation}\label{eq:Vnm1}\begin{split}
  V_{n-1}(x,t) \equiv V_{n-1}^{\rm f}(x,t) = \sum_{p=1}^{r} \frac{\alpha_p(K) }{\alpha_p(J)} q_p(x,t) \mathcal{E}_p^{(n_1-1)}.
 \end{split}\end{equation}
\end{remark}

Now let us assume that $s_1$ is an exponent and split the third equation in (\ref{Recurrence1}) into diagonal and off-diagonal parts.
Evaluating the Killing form of this equation with $\mathcal{H}_1^{h-s_1}$ we obtain:
\begin{equation}\label{eq:ws}\begin{split}
 w_s(x,t) = \frac{i}{c_{s_1}} \partial_x^{-1} \left\langle [Q,V_s^{\rm f}] ,\mathcal{H}_1^{h-s_1} \right\rangle + \const, \qquad
 c_{s_1} = \left\langle \mathcal{H}_1^{s_1}, \mathcal{H}_1^{h-s_1}\right\rangle .
\end{split}\end{equation}
In what follows for simplicity we will set all these integration constants to 0. A diligent reader can easily work out the more
general cases when some of these constants do not vanish.
The off-diagonal part of the third equation in (\ref{Recurrence1}) gives:
\begin{equation}\label{eq:Vsm1f}\begin{split}
i \partial_x V_s^{\rm f} + [Q, V_s^{\rm f}]^{\rm f} + [Q, w_s \mathcal{H}_1^{s_1}] = [J, V_{s-1}].
\end{split}\end{equation}
i.e.
\begin{equation}\label{eq:Vsm1f2}\begin{split}
 V_{s-1}^{\rm f} = \ad_J^{-1} \left( i \partial_x V_s^{\rm f} + [Q, V_s^{\rm f}]^{\rm f} + [Q, w_s \mathcal{H}_1^{s_1}]\right) = \Lambda_{s_1} V_s^{\rm f}.
\end{split}\end{equation}
Thus we obtained the integro-differential operator $\Lambda_{s_1} $ which acts on any $Z \equiv Z^{\rm f} \in \mathfrak{g}^{(s_1)}$ by:
\begin{equation}\label{eq:Lams1}\begin{split}
 \Lambda_{s_1} Z = \ad_J^{-1} \left( i \partial_x Z + [Q, Z]^{\rm f} + \frac{i}{c_{s_1}}[Q,  \mathcal{H}_1^{s_1}]
 \partial_x^{-1} \left\langle [Q,Z] ,\mathcal{H}_1^{h-s_1} \right\rangle  \right)
\end{split}\end{equation}
If $s_1$ is not an exponent we have only to work out the off-diagonal part of the third equation in (\ref{Recurrence1})
with the result:
\begin{equation}\label{eq:Vsm1f3}\begin{split}
 V_{s-1}^{\rm f} &= \ad_J^{-1} \left( i \partial_x V_s^{\rm f} + [Q, V_s^{\rm f}]^{\rm f} \right) = \Lambda_{0} V_s^{\rm f}, \\
 \Lambda_0 Z &= \ad_J^{-1} \left( i \partial_x Z + [Q, Z]^{\rm f} \right).
\end{split}\end{equation}
Now $\Lambda_0$ is a differential operator.

In the case of $D_4^{(1)}$ the exponent 3 has multiplicity 2. Therefore the recursion operator $\Lambda_3$ must be replaced
by $\tilde{\Lambda}_3$ which has the form:
\begin{equation}\label{eq:La3t}\begin{split}
\tilde{\Lambda}_{3} Z = \ad_J^{-1} \left( i \partial_x Z + [Q, Z]^{\rm f} + \frac{i}{c_3}[Q,  \mathcal{H}_1^{3}]
 \partial_x^{-1} \left\langle [Q,Z] ,\mathcal{H}_1^{3} \right\rangle +\frac{i}{c_3'}[Q,  \mathcal{H}_4^{3}]
 \partial_x^{-1} \left\langle [Q,Z] ,\mathcal{H}_4^{3} \right\rangle \right)
\end{split}\end{equation}
where
\begin{equation}\label{eq:c3}\begin{split}
  c_{3} = \left\langle \mathcal{H}_1^{3}, \mathcal{H}_1^{3}\right\rangle , \qquad   c_{3}' = \left\langle \mathcal{H}_4^{3}, \mathcal{H}_4^{3}\right\rangle .
\end{split}\end{equation}
Here we also used the fact that $ \left\langle \mathcal{H}_1^{3}, \mathcal{H}_4^{3}\right\rangle =0$.

\subsection{The hierarchies of MKdV related to $D_4^{(a)}$}
The last of the equations in (\ref{Recurrence1}) provides the corresponding set of MKdV equations that can be solved
applying the ISM to the corresponding Lax operator. In fact this last equations simplifies into
\begin{equation}
\label{Hierarchy}
{\partial_t Q(x,t)} = \partial_x V_0(x,t).
\end{equation}
because the subalgebra $\mathfrak{g}^{(0)}$ is commutative. Let us now describe the class of the sets of MKdV
equations using the recursion operators $\Lambda_0$ and $\Lambda_s$.
With each of the Lax operators $L$ we can relate 4 series of NLEE whose dispersion laws are monomial in $\lambda$.

Let us first start with $D_4^{(1)}$. Skipping the details we write them compactly as follows:
\begin{equation}\label{eq:D41Mkdv}\begin{aligned}
n&= 6n_0 +1 &\quad \partial_t Q &= \partial_x \left(\bm{\Lambda}^{n_0} Q(x,t)\right) ,&\quad  f(\lambda)&= \lambda^{6n_0+1} \mathcal{H}_1^{(1)}, \\
n&= 6n_0 +3 &\quad \partial_t Q &= \partial_x \left( \bm{\Lambda}^{n_0} \Lambda_1 \Lambda_0 \ad_J^{-1} [a\mathcal{H}_1^{3} +b\mathcal{H}_4^{3}, Q(x,t)] \right),
&\; f(\lambda) &= \lambda^{6n_0+3}(a\mathcal{H}_1^{3} +v\mathcal{H}_4^{3}), \\
n&= 6n_0 +5 &\quad \partial_t Q &= \partial_x \left(\bm{\Lambda}^{n_0} \Lambda_1 \Lambda_0 \tilde{\Lambda}_3 \Lambda_0
 \ad_J^{-1} [\mathcal{H}_1^{5}, Q(x,t)]\right) , &\quad
f(\lambda) &= \lambda^{6n_0+5} \mathcal{H}_1^{5}.
\end{aligned}\end{equation}
where $\bm{\Lambda}=\Lambda_1 \Lambda_0 \tilde{\Lambda}_3\Lambda_0 \Lambda_5\Lambda_0$.
The fact that the exponent 3 is double valued leads to the fact that with each dispersion law proportional to $\lambda^{6n_0+3}$ we have
a one-parameter family of NLEE. Indeed, we can rescale the time $t \to \tau = t/a$ which will make  the parameter $a=1$; however the other parameter
$b \to b/a$ can not be taken away.

Similarly we can treat the hierarchies related to $D_4^{(2)}$. The results are:
\begin{equation}\label{eq:D42Mkdv}\begin{aligned}
n&= 8n_0 +1 &\quad \partial_t Q &= \partial_x \left(\bm{\Lambda}^{n_0} Q(x,t)\right) ,&\quad  f(\lambda) &= \lambda^{8n_0+1} \mathcal{H}_1^{(1)}, \\
n&= 8n_0 +3 &\quad \partial_t Q &= \partial_x \left( \bm{\Lambda}^{n_0} \Lambda_1 \Lambda_0 \ad_J^{-1} [\mathcal{H}_1^{3}, Q(x,t)] \right),
&\quad f(\lambda) &= \lambda^{8n_0+3}\mathcal{H}_1^{3}, \\
n&= 8n_0 +5 &\quad \partial_t Q &= \partial_x \left(\bm{\Lambda}^{n_0} \Lambda_1 \Lambda_0 \Lambda_3 \Lambda_0  \ad_J^{-1}[\mathcal{H}_1^{5}, Q(x,t)]\right) , &\quad
f(\lambda) &= \lambda^{8n_0+5} \mathcal{H}_1^{5}, \\
n&= 8n_0 +7 &\quad \partial_t Q &= \partial_x \left(\bm{\Lambda}^{n_0} \Lambda_1 \Lambda_0 \Lambda_3 \Lambda_0 \Lambda_5 \Lambda_0
 \ad_J^{-1}[\mathcal{H}_1^{7}, Q(x,t)]\right) , &\quad
f(\lambda) &= \lambda^{8n_0+7} \mathcal{H}_1^{7},
\end{aligned}\end{equation}
where $\bm{\Lambda}=\Lambda_1 \Lambda_0 \Lambda_3\Lambda_0 \Lambda_5\Lambda_0 \Lambda_7\Lambda_0$.

Finally for $D_4^{(3)}$ we get:
\begin{equation}\label{eq:D43Mkdv}\begin{aligned}
n&= 12n_0 +1 &\quad \partial_t Q &= \partial_x \left(\bm{\Lambda}^{n_0} Q(x,t)\right) ,&\quad  f(\lambda) &= \lambda^{12n_0+1} \mathcal{H}_1^{(1)}, \\
n&= 12n_0 +5 &\quad \partial_t Q &= \partial_x \left( \bm{\Lambda}^{n_0} \Lambda_1 \Lambda_0^3  \ad_J^{-1} [\mathcal{H}_1^{5}, Q(x,t)] \right),
&\quad f(\lambda) &= \lambda^{12n_0+5}\mathcal{H}_1^{5}, \\
n&= 12n_0 +7 &\quad \partial_t Q &= \partial_x \left(\bm{\Lambda}^{n_0} \Lambda_1 \Lambda_0^3 \Lambda_5 \Lambda_0  \ad_J^{-1} [\mathcal{H}_1^{7}, Q(x,t)]\right) , &\quad
f(\lambda) &= \lambda^{12n_0+7} \mathcal{H}_1^{7}, \\
n&= 12n_0 +11 &\quad \partial_t Q &= \partial_x \left(\bm{\Lambda}^{n_0} \Lambda_1 \Lambda_0^3 \Lambda_5 \Lambda_0 \Lambda_7 \Lambda_0^3
 \ad_J^{-1} [\mathcal{H}_1^{11}, Q(x,t)]\right) , &\quad f(\lambda) &= \lambda^{12n_0+11} \mathcal{H}_1^{11},
\end{aligned}\end{equation}
where $\bm{\Lambda}=\Lambda_1 \Lambda_0^3 \Lambda_5\Lambda_0 \Lambda_7\Lambda_0^3 \Lambda_{11}\Lambda_0$.

We end this Section by the simple remark, which follows directly from the structure of the recursion operators and from
the grading conditions of the algebras. Indeed, since $J \in \mathfrak{g}^{(1)}$ and $Q \in \mathfrak{g}^{(0)}$, then
\begin{equation}\label{eq:adJ}\begin{aligned}
 \ad_J &\colon  \mathfrak{g}^{(p)} \to \mathfrak{g}^{(p+1)}, &\qquad  \ad_J^{-1} &\colon  \mathfrak{g}^{(p)} \to \mathfrak{g}^{(p-1)}, &\qquad
  \ad_Q &\colon  \mathfrak{g}^{(p)} \to \mathfrak{g}^{(p)}, \\
\Lambda_{k} &\colon  \mathfrak{g}^{(p)} \to \mathfrak{g}^{(p-1)}, &\qquad \Lambda_{k} &\colon  \mathfrak{g}^{(p)} \to \mathfrak{g}^{(p-1)}, &\qquad
\bm{\Lambda} &\colon  \mathfrak{g}^{(p)} \to \mathfrak{g}^{(p)}.
\end{aligned}\end{equation}
Thus it is easy to check that both sides of the NLEE (\ref{eq:D41Mkdv}),  (\ref{eq:D42Mkdv}) and  (\ref{eq:D43Mkdv}) obviously take values
in $ \mathfrak{g}^{(0)}$.



\section{Hamiltonian formulation and the first non-trivial members of the hierarchies}

The first non-trivial member of the hierarchy is a set of mKdV equations which is obtained from (\ref{eq:D41Mkdv}),  (\ref{eq:D42Mkdv}) and  (\ref{eq:D43Mkdv})
setting $n=3$, $n=3$ and $n=5$ respectively.

Here we will briefly describe the Hamiltonian formulation of the equations from \eqref{Hierarchy}.
Every equation in \eqref{Hierarchy} has infinitely many integrals of motion
Every integral of motion can be viewed as a Hamiltonian with a properly chosen Poisson structure.
We will use the integral of motion given by \cite{VG-Ya-13, VG-Ya-14}
\begin{equation}
I = \int_{{-\infty}}^{\infty} i \partial_x^{-1} \left< \left[ Q,  \Lambda_{0} V^{(0)} \right], \mathcal{H}^{(1)}_1 \right>{dx},
\label{I1}
\end{equation}
where $\partial_x^{-1} f(x) = \int f(x) dx$ and we have set any constants of integration to be zero.
The Hamiltonian $H$ is proportional to \eqref{I1}.
Hamilton's equations are
\begin{equation}
\partial_{t} q_i = \{ q_i, H \}
\label{HamEq}
\end{equation}
with a Poisson bracket given by
\begin{equation}
\{F,G\} = \int_{\mathbb{R}^2} \omega_{ij}(x,y) \frac{\delta{F}}{\delta{q_i}}\frac{\delta{G}}{\delta{q_j}} {dx} {dy},
\end{equation}
where we sum over repeating indexes.
The Poisson structure tensor is
\begin{equation}
\omega_{ij}(x,y)= \frac{1}{2} \delta_{ij} \left( \partial_{x} \delta(x-y) - \partial_{y} \delta(x-y) \right).
\end{equation}
In this case \eqref{HamEq} reduces to
\begin{equation}
\partial_t q_i = \partial_x \frac{\delta{H}}{\delta{q_i}}.
\label{HamEq2}
\end{equation}

\subsection{The MKdV for $D_4^{(1)}$ }
The algebra $D_4^{(1)}$ has 3 as a double exponent. This means that the
element $K$ in \eqref{LaxPair} involves two arbitrary parameters:
\begin{equation}
K = \frac{1}{2} a \mathcal{H}^{(3)}_1 + \frac{1}{6}b \mathcal{H}^{(3)}_4.
\end{equation}
We can also say, that there are two nonequivalent mKdV equations related to $D_4^{(1)}$. Each equation  is determined by
its own dispersion law: the first one by $f_1 = \frac{1}{2}  \lambda^3 \mathcal{H}^{(3)}_1 $, the other one -- by
$f_2 = \frac{1}{2}  \lambda^3 \mathcal{H}^{(3)}_4 $. Below we will write down the two systems of equations separately;
the generic mKdV equation will have as dispersion law $\lambda^3 K$ which is a linear combination of $f_1$ and $f_2$.
\begin{equation}\label{eq:mkdv1a}\begin{aligned}
\partial_t q_1 &= \partial_x \left( 2\partial^2_x q_1 - \sqrt{3} \left( 2  q_1 \partial_x q_2 +3q_4 \partial_x q_3 +3q_3 \partial_x q_4 \right)
+ 3 (-2  q_2^2 +q_3^2 +q_4^2)q_1 \right),\\
\partial_t q_2 &= \partial_x \left(  \sqrt{3} \left( 2q_1 \partial_x q_1 -q_3 \partial_x q_3 - q_4 \partial_x q_4 \right) +3( -2q_1^2+ q_3^2 +  q_4^2)q_2 \right)   , \\
\partial_t q_3 &= \partial_x \left( - \partial^2_x q_3 + \sqrt{3} \left( q_3 \partial_x q_2 + 3q_4 \partial_x q_1\right) +
3  \left(q_1^2 + q_2^2 -2 q_4^2\right) q_3\right) ,\\
\partial_t q_4 &= \partial_x \left(  - \partial^2_x q_4 + \sqrt{3} \left( 3q_3 \partial_x q_1 + q_4 \partial_x q_2\right) +
3  \left(q_1^2 + q_2^2 -2 q_3^2\right) q_4\right) .
\end{aligned}
\end{equation}
The second set of mKdV eqs. are given by:
\begin{equation}\label{eq:mkdv1b}\begin{aligned}
\partial_t q_1 &= \partial_x \left( \sqrt{3} \left(   q_3 \partial_x q_4 - q_4 \partial_x q_3\right) + 3 (q_4^2 -q_3^2)q_1 \right),\\
\partial_t q_2 &= \partial_x \left( -\sqrt{3} \left(   q_3 \partial_x q_3 - q_4 \partial_x q_4\right) +3 (q_3^2 -q_4^2)q_2  \right)  , \\
\partial_t q_3 &= \partial_x \left( - \partial^2_x q_3 + \sqrt{3} \left( 2 q_1 \partial_x q_4 +q_3 \partial_x q_2+ q_4 \partial_x q_1\right) +
3  \left(q_2^2 - q_1^2\right) q_3\right) ,\\
\partial_t q_4 &= \partial_x \left(  \partial^2_x q_4 - \sqrt{3} \left( q_3 \partial_x q_1 + q_4 \partial_x q_2 + 2q_1 \partial_x q_3\right) +
3  \left(q_1^2 - q_2^2 \right) q_4\right) .
\end{aligned}
\end{equation}

The Hamiltonian densities of these equations are given by:
\begin{equation}\label{eq:mkdv1Ha}\begin{split}
 H_a &= -(\partial_x q_1)^2 + \frac{1}{2} (\partial_x q_3)^2 + \frac{1}{2} (\partial_x q_4)^2  + 3\sqrt{3} q_3q_4 \partial_x q_1 - \frac{\sqrt{3}}{2}
 (2q_1^2 -q_3^2 - q_4^2) \partial_x q_2 \\
  &+ \frac{3}{2} (q_1^2 +q_2^2)(q_3^2 +q_4^2) - 3q_1^2q_2^2 - 3q_3^2q_4^2 .
\end{split}\end{equation}
and
\begin{equation}\label{eq:mkdv1Hb}\begin{split}
 H_b =  \frac{1}{2} (\partial_x q_3)^2 - \frac{1}{2} (\partial_x q_4)^2  + \frac{ \sqrt{3}}{2} \left( (q_3^2 - q_4^2) \partial_x q_2
 - 2 q_1q_4 \partial_x q_3 + 2 q_1q_3 \partial_x q_4\right) + \frac{3}{2} (q_4^2 -q_3^2)(q_1^2 - q_2^2) .
\end{split}\end{equation}

Notice that the equation for $q_2$ in (\ref{eq:mkdv1a}), as well as the equations for $q_1$ and $q_2$ in (\ref{eq:mkdv1b}) do not contain
third order derivatives with respect to $x$. This fact  directly reflects the structure of the Hamiltonians. Indeed, in the kinetic past of $H_a$
there is no term proportional to $(\partial_x q_2)^2 $. Likewise the kinetic part of $H_a$ there are no term proportional to $(\partial_x q_1)^2 $
and $(\partial_x q_2)^2 $. Another important property of these Hamiltonians is that their kinetic parts are neither positive nor negative definite.

\subsection{The MKdV for $D_4^{(2)}$  }
Since the rank of $D_4^{(2)}$ is 3 then the mKdV  is a set of three equations for three functions. Since the 3 is an exponent,
then the simplest mKdV will contain third order derivatives with respect to $x$.  The details of the calculations and the explicit form
of these equations can be found in \cite{GMSV4}.

\begin{equation}\label{eq:mkd42}\begin{split}
\partial_t q_1 & = 2\partial_x\Big((4+3\sqrt2)\partial_x^2q_1-3(2+\sqrt2)q_1\partial_x q_2
                 -6\sqrt2q_2\partial_x q_3-3\sqrt2 q_3\partial_x q_2\\
               & \qquad  +2q_3(3q_1^2-6q_2^2+q_3^2-6q_1q_3)-6q_1q_2^2\Big),\\
\partial_t q_2 & =2\partial_x \Big(\partial_x^2 q_2+3(2+\sqrt2)q_1\partial_x q_1+3(2-\sqrt2)q_3\partial_x q_3
                     +3\sqrt2 q_3\partial_x q_1-3\sqrt2 q_1\partial_x q_3\\
               & \qquad  - 2q_2 (3q_1^2+q_2^2+3q_3^2+12q_1q_3)\Big),\\
\partial_t q_3 & =2\partial_x\Big((4-3\sqrt2)\partial_x^2q_3-3(2-\sqrt2)q_3\partial_x q_2
                        +6\sqrt2 q_2\partial_x q_1+3\sqrt2 q_1\partial_x q_2\\
               & \qquad +2q_1(q_1^2-6q_2^2+3q_3^2-6q_1q_3)-6q_2^2q_3\Big).
\end{split}\end{equation}

The Hamiltonian density of this system is:
\begin{equation}\label{eq:HD42}\begin{split}
 H &= -(4+3\sqrt{2}) (\partial_x q_1)^2 - (\partial_x q_2)^2 -(4-3\sqrt{2}) (\partial_x q_3)^2 -3((2+\sqrt{2})q_1^2 +(2-\sqrt{2})q_3^2) \partial_x q_2 \\
& - 6\sqrt{2} q_2\left(q_1 \partial_x q_3 - q_3 \partial_x q_1 \right)
+ 4q_1q_3 (q_1^2 + q_3^2-3q_1q_3) -  q_2^2 (6q_1^2 + q_2^2+6q_3^2+24q_1q_3).
\end{split}\end{equation}

Like in the previous cases, the kinetic part of $H^{(2)}$ is not positive definite. However now it contains all three fields
$q_i$, so each of the equations will contain terms with third order derivatives with respect to $x$.

\subsection{The MKdV for $D_4^{(3)}$ }

The rank of $D_4^{(3)}$ is 2 then the mKdV  is a set of two equations for two functions. Now the set of exponents is 1, 5, 7 and 11.
Therefore the simplest mKdV equations will be a set of two equations of fifth order with respect to $x$.
Skipping the details we  write down the equations with dispersion law $f_5 = \lambda^5 \mathcal{H}_1^{(5)}$:
\begin{multline}\label{eq:qt1}
\partial_t q_1 =  \frac{\partial }{ \partial x }\Big( (3 \sqrt{3}+5) \partial_x^4 q_1  +10(q_1-q_2) \partial_x^3 q_2 +
5 \left( (5\sqrt{3}+9) \partial_x q_1 +(\sqrt{3}+1) \partial_x q_2 \right) \partial_x^2 q_1 \\
 -20 \left( (\sqrt{3}+1) (q_1^2 +q_2^2)+q_1q_2 \right) \partial_x^2 q_1 +5 \left( (\sqrt{3}+1) \partial_x q_1 -(\sqrt{3}+5) \partial_x q_2 \right) \partial_x^2 q_2 \\
 -10 \left( (\sqrt{3}+3) q_1^2 -(3-\sqrt{3})q_2^2  \right) \partial_x^2 q_2 -10 \left( 2(\sqrt{3}+1)q_1 +q_2 \right) (\partial_x q_1)^2 \\
 -20 \left( q_1 + 2(\sqrt{3}+1) q_2 \right) \partial_x q_1  \partial_x q_2 +10 \left( 2(\sqrt{3}-1)q_1 +(5-2\sqrt{3})q_2 \right) (\partial_x q_2)^2 \\
+ 20(q_1-q_2) \left( (2\sqrt{3}+1) q_1^2 +4q_1q_2 -(2\sqrt{3}-1)q_2^2 \right) \partial_x q_2 \\
+ 10\sqrt{3} \left( q_1^5  -4q_1^4q_2 +2q_1^3q_2^2 +8 q_1^2q_2^3 +q_1q_2^4 \right) - 8 \sqrt{3} q_2^5 \Big).
\end{multline}
\begin{multline}\label{eq:qt2}
\partial_t q_2 = \frac{\partial }{ \partial x } \Big( (3 \sqrt{3}-5) \partial_x^4 q_2  -10(q_1-q_2) \partial_x^3 q_1 -
5 \left( (5-\sqrt{3}) \partial_x q_1 +(\sqrt{3}-1) \partial_x q_2 \right) \partial_x^2 q_1 \\
 -10 \left( (3+\sqrt{3}) q_1^2 -  (3-\sqrt{3}) q_2^2) \right) \partial_x^2 q_1 +5 \left( -(\sqrt{3}-1) \partial_x q_1 +(9- 5\sqrt{3}) \partial_x q_2 \right) \partial_x^2 q_2 \\
 +20 \left( -(\sqrt{3}-1) (q_1^2 +q_2^2) +q_1q_2  \right) \partial_x^2 q_2 -10 \left( (2\sqrt{3}+5)q_1 - 2(\sqrt{3}+1)q_2 \right) (\partial_x q_1)^2 \\
 -20 \left( 2(\sqrt{3}-1)q_1 -  q_2 \right) \partial_x q_1  \partial_x q_2 +10 \left(  q_1 -2(\sqrt{3}-1)q_2 \right) (\partial_x q_2)^2 \\
- 20(q_1-q_2) \left( (2\sqrt{3}+1) q_1^2 +4q_1q_2 -(2\sqrt{3}-1)q_2^2 \right) \partial_x q_1 \\
+   10\sqrt{3} q_2(q_1^2 +q_2^2)^2 -8\sqrt{3} q_1(q_1^4 -10 q_1^2q_2^2+5q_2^4)    \Big).
\end{multline}
\begin{multline}\label{eq:HD3}
 H_3= \frac12\left[ (3\sqrt3+5)(\partial_x^2q_1)^2+(3\sqrt3-5)(\partial_x^2q_2)^2+5(q_2^2-2q_1q_2)\partial_x^3q_1
            +5(q_1^2-2q_1q_2)\partial_x^3q_2\right] \\
           +\frac52\left[ (2+\sqrt3)\partial_xq_1(\partial_xq_2)^2-(3+\frac53\sqrt3)(\partial_xq_1)^3
                        -(3-\frac53\sqrt3)(\partial_xq_2)^3+(2-\sqrt3)(\partial_xq_1)^2\partial_xq_2\right] \\
              +10\left[ (\sqrt3+1) (q_1^2 +q_2^2)+q_1q_2 \right] (\partial_x q_1)^2
               +10\left[ (\sqrt3-1)(q_1^2+q_2^2) -q_1q_2 \right] (\partial_x q_2)^2\\
            +4(q_2-q_1)\left[ (1+2\sqrt3)q_1^2+4q_1q_2+(1-2\sqrt3)q_2^2\right] (q_2\partial_xq_1-q_1\partial_xq_2)\\
                   +10\left[ (\sqrt3+3)q_1^2+(\sqrt3-3)q_2^2\right] \partial_xq_1\partial_xq_2  + \frac{\sqrt3}{3}\left[ 5(q_1^2+q_2^2)^3-24(q_1^4+q_2^4)q_1q_2+80q_1^3q_2^3\right]
\end{multline}

We note that the equations (\ref{eq:qt1}) and (\ref{eq:qt2}) can be written down in the form:
\begin{equation}\label{eq:Ili1}\begin{split}
 \partial_t q_1 &=  \partial_x (P(q_1,q_2)+Q(q_1,q_2)) \\
\partial_t q_2 &=   \partial_x (P(q_2,q_1)-Q(q_2,q_1))
\end{split}\end{equation}
where 
\begin{equation}\label{eq:Ili2}\begin{split}
P(u,v) &= 3\sqrt3\partial^4_xu+10(u-v)\partial^3_xv+45\partial_x u\partial_x^2u+5\partial_x v\partial_x^2u
              +5\partial_x u\partial_x^2 v-25\partial_x v\partial_x^2v\\
            &-20\sqrt3(u^2+v^2)\partial^2_x u -10\sqrt3(u^2+v^2)\partial_x^2 v
           -20\sqrt3 u(\partial_xu)^2+20\sqrt3 u(\partial_x v)^2\\
            &-20\sqrt3 v(\partial_x v)^2-40\sqrt3 v\partial_x u \partial_x v+20(u^3+3u^2v-3uv^2-v^3)\partial_x v\\
             & +2\sqrt3 (5u^5-20u^4v+10u^3v^2+40u^2v^3+5uv^4-4v^5),\\
Q(u,v) &=  5 \partial^4_x u+25\sqrt3 \partial_x u\partial_x^2u+5\sqrt3\partial_x v\partial_x^2u
                                 +5\sqrt3 \partial_x u\partial_x^2 v-5\sqrt3 \partial_x v\partial_x^2 v\\
             & -20(u^2+uv+v^2)\partial^2_x u -30(u^2-v^2)\partial^2_x v-20u(\partial_x u)^2 -10v(\partial_x u)^2
               -20u(\partial_x v)^2\\
             &+50 v(\partial_x v)^2
              -20 u\partial_x u\partial_x v-40 v\partial_x u\partial_x v+40\sqrt3(u^3-u^2v-uv^2+v^3)\partial_xv.
\end{split}\end{equation}

\section{Discussion and conclusions}
We constructed the three nonequivalent Coxeter gradings in the algebra $D_4 \simeq so(8)$. The first of them is the
standard one obtained with the Coxeter automorphism $C_1=S_{\alpha_2} S_{\alpha_1}S_{\alpha_3}S_{\alpha_4}$ using
its dihedral realization. In the second one we use $C_2 = C_1R$ where $R$ is the mirror automorphism. The third one
is $C_3 = S_{\alpha_2}S_{\alpha_1}T$ where $T$ is the external automorphism of order 3. For each of these gradings we
constructed the basis in the corresponding linear subspaces $\mathfrak{g}^{(k)}$, the orbits of the Coxeter automorphisms
and the related Lax pairs generating the corresponding mKdV hierarchies. We found compact expressions for each of the hierarchies
in terms of the recursion operators. At the end we wrote explicitly the first nontrivial mKdV equations and their Hamiltonians.
For $D_4^{(1)}$ these are in fact two mKdV systems, due to the fact that in this case the exponent $3$ has multiplicity 2.
Each of these mKdV systems consist of 4 equations of third order with respect to $\partial_x$. For $D_4^{(2)}$ this is a
system of three equations of third order with respect to $\partial_x$. Finally, for $D_4^{(3)}$ this is a system of two
equations of fifth order with respect to $\partial_x$.

The fact that these mKdV equations have the structure outlined above is a consequence of the seminal papers by
Mikhailov \cite{Mikhailov} and Drinfeld and Sokolov \cite{DriSok}. However the explicit formulation of the Lax operators
as well as the explicit form of the equations themselves and their Hamiltonians, especially the
ones for $D_4^{(2)}$ and $D_4^{(3)}$ are not so well known and deserve additional studies.  Indeed, the fact that
the mKdV equations (\ref{eq:qt1}) and (\ref{eq:qt2}) can be cast into the form (\ref{eq:Ili1}), (\ref{eq:Ili2}) means,
that the Lax pair for  $D_4^{(3)}$ case has a symmetry that interchanges the $q_1 \leftrightarrow q_2$ combined with
a Weyl reflection $S_{e_1-e_3} S_{e_2+e_4} $ which interchanges the orbits $\mathcal{O}_1 \leftrightarrow \mathcal{O}_2$.
Similar more complicated symmetries exist also for $D_4^{(1)}$ and $D_4^{(2)}$; they will be studied in next publications.

We note, that the problem of constructing  systems of higher mKdV equations for two functions has been attacked by
using the symmetry formalism developed by Shabat and his collaborators, see \cite{MiShaYa} and the references therein.
In the \cite{MNW,Novik1} a system of two mKdV equations of order 5 with exponents 1, 5, 7 and 11 has been reported.
It does not coincide with the system (\ref{eq:qt1}) and (\ref{eq:qt2}) found above. On the other hand the only Kac-Moody
algebra that has rank 2 and exponents 1, 5, 7 and 11 is $D_4^{(3)}$, so the two systems must be equivalent. In other
words one should be looking for a (gauge) transformation that relates the two equations.

Another important aspect in the studies of these equations is related to the spectral theory of the corresponding Lax
operators. This will require further elaboration of the results in \cite{VG-Ya-13, VG-Ya-14} specifying them to the relevant choices of $Q(x,t)$
and $J$ in the Lax operators. One can expect deeper understanding of the expansions over the `squared` solutions of $L$.
As a result one could see that even for Lax operators possessing deep reductions the inverse scattering problem can be
related to a Riemann-Hilbert problem, and can be interpreted as a generalized Fourier transform \cite{VG-Ya-13, VG-Ya-14}. In particular one can expect
to derive the symplectic form of the `squared solutions' \cite{GeKh1, IlKh} and as a result to derive
explicit expressions for the action-angle variables for the mKdV hierarchy in terms of the scattering data. All these results will
be naturally compatible with the existence of the hierarchy of Hamiltonian structures of the mKdV equations \cite{DriSok, FaTa, ContM, Basic, VG-Ya-14, GeYaV}
and the hierarchy of Lagrangian structures \cite{NutPav}.

\section*{Acknowledgements}
 We are grateful to Ms S. Sushko for careful reading of the manuscript.
One of us (VSG) is grateful to professor A. V. Mikhailov and professor V. S. Novikov for useful
discussions and comments.
This work has been supported by the Bulgarian Science Foundation (grant NTS-Russia 02/101
from 23.10.2017) and by  the RFBR  (grant 18-51-18007).
Two of us (VSG and AAS) are grateful to the organizing committee of the
IX-th International Conference (SCT-19) ``Solitons, collapses and turbulence''
Achievements, Developments and Perspectives  in honor of Vladimir Zakharov's 80th birthday, held in Yaroslavl, Russia August 5-9, 2019 for their support and hospitality.




\end{document}